\documentclass[preprint]{aastex}
\usepackage{natbib}
\usepackage{amsfonts}

\slugcomment{}

\shorttitle{Magnetic Field Structure around Low-Mass Class 0 Protostars}
\shortauthors{Davidson et al.}

\begin{document}

\title{Magnetic Field Structure around Low-Mass Class 0 Protostars: B335, L1527 and IC348-SMM2}

\author{J.A. Davidson\altaffilmark{1}}
\affil{University of Western Australia} 

\author{G. Novak\altaffilmark{2} and T. G. Matthews\altaffilmark{2}}
\affil{Northwestern University}

\author{B. Matthews\altaffilmark{3}$^,$\altaffilmark{4} }
\affil{Herzberg Institute of Astrophysics \& University of Victoria}

\author{P. F. Goldsmith and N. Chapman\altaffilmark{5}}
\affil{Jet Propulsion Laboratory, California Institute of Technology}

\author{N. H. Volgenau\altaffilmark{6}}
\affil{California Institute of Technology} 

\author{J. E. Vaillancourt\altaffilmark{7}}
\affil{Universities Space Research Association}

\and

\author{M. Attard\altaffilmark{8}}
\affil{Service d'Astrophysique, CEA Saclay}

\altaffiltext{1}{University of Western Australia, School of Physics, 35 Stirling Hwy, Crawley, WA 6009, Australia}
\altaffiltext{2}{Northwestern University, Department of Physics and Astronomy, 2131 Tech Dr., Evanston, IL 60208, USA} 
\altaffiltext{3}{Herzberg Institute of Astrophysics, National Research Council of Canada, 5071 West Saanich Road, Victoria, BC, V9E 2E7, Canada}
\altaffiltext{4}{University of Victoria, 3800 Finnerty Road, Victoria, BC, V8P 1A1, Canada} 
\altaffiltext{5}{Jet Propulsion Laboratory, California Institute of Technology, 4800 Oak Grove Dr., Ms 264-782, Pasadena, CA 91109, USA}
\altaffiltext{6}{California Institute of Technology, Owens Valley Radio Observatory, Big Pine, CA 93513, USA}
\altaffiltext{7}{Universities Space Research Association, SOFIA, NASA Ames Research Center, MS 211-3, Moffett Field, CA 94035-0001, USA}
\altaffiltext{8}{Laboratoire AIM, CEA/DSM-CNRS-Universit\'{e} Paris Dierot, 91191 Gif-sur-Yvette, France}

\begin{abstract}

We report new 350 $\mu$m polarization observations of the thermal dust emission from the cores surrounding the low-mass, Class 0 YSOs L1527, IC348-SMM2 and B335.  We have inferred magnetic field directions from these observations, and have used them together with results in the literature to determine whether magnetically regulated core-collapse and star-formation models are consistent with the observations.  These models predict a pseudo-disk with its symmetry axis aligned with the core magnetic field.  The models also predict a magnetic field pinch structure on a scale less than or comparable to the infall radii for these sources.  In addition, if the core magnetic field aligns (or nearly aligns) the core rotation axis with the magnetic field before core collapse, then the models predict the alignment (or near alignment) of the overall pinch field structure with the bipolar outflows in these sources.
We show that if one includes the distorting effects of bipolar outflows on magnetic fields, then in general the observational results for L1527 and IC348-SMM2 are consistent with these magnetically regulated models.  We can say the same for B335 only if we assume the distorting effects of the bipolar outflow on the magnetic fields within the B335 core are much greater than for L1527 and IC348-SMM2.  
We show that the energy densities of the outflows in all three sources are large enough to distort the magnetic fields predicted by magnetically regulated models.

\end{abstract}


\section{Introduction}

Only a few percent of the mass of a molecular cloud is typically converted into stars (McKee \& Ostriker 2007; Goldsmith et al. 2008).  The rate of star formation inferred for molecular clouds in general is significantly less than that expected from free-fall gravitational collapse of the gravitationally bound mass in these regions.  Consequently, there must be some mechanism which regulates the rate of star formation within molecular clouds. Two mechanisms have been proposed; magnetic support (e.g., Shu, Adams \& Lizano 1987; Mouschovias \& Ciolek 1999) and super-Alfvenic  turbulence (e.g., Mac Low \& Klessen 2004).  In addition, observations (Goldsmith \& Arquilla 1985; Goodman et al. 1993) show that the rotation rates of cores (density $\gtrsim$ 10$^4$ cm$^{-3}$; size $\sim$ 0.1 pc) are less than expected for cores condensed from their less dense background clouds.   In the models favoring magnetic support, this slow core rotation is the result of magnetic braking (Mestel 1985; Basu \&  Mouschovias 1994).  Mouschovias \& Paleologou (1979, 1980) showed analytically that the braking timescale for a cloud rotating with its rotation axis perpendicular to the magnetic field is typically an order of magnitude smaller than for a cloud rotating with its axis of rotation parallel to the magnetic field. This results in core rotation rates reduced in amplitude and core rotation axes tending to align with the cloud magnetic field immediately in the vicinity of cores.   In this paper, we use our submillimeter polarimetry results 
to test gross features of
the magnetically regulated core-collapse and star-formation models in the literature.

A key feature of many such magnetically regulated dynamical collapse models (Shu, Adams \& Lizano 1987; Allen, Shu \& Li 2003; Allen, Li \& Shu 2003; Galli \& Shu 1993a,b; Shu, Li \& Allen 2004; Tomisaka 1998) is the formation of a flattened inner core (or ``pseudo-disk'') of several thousand AU in extent with its symmetry axis parallel to the core magnetic field. This pseudo-disk is the inner part of the infall region of the core surrounding a YSO; it is not formed by rotation, but by the geometry of the magnetic field.  The pseudo-disk is a dynamically collapsing entity, accreting onto the protostar and its associated Keplerian disk (size $\sim$ 100 AU).  Signatures of magnetic regulation include a fairly uniform magnetic field outside the infall region of a core, a pinched magnetic field line structure within this region, and the overall direction of the magnetic field parallel to the axis of symmetry of the pseudo-disk. The level of uniformity of the field outside the infall region depends on the quasi-static evolution prior to the dynamical collapse of the core.  In the Galli \& Shu (1993a,b) models, the field lines outside the infall region are uniform, but in the Allen, Shu \& Li (2003; hereafter ASL03a) and Allen, Li \& Shu (2003; hereafter ALS03b) models, they are already inclined towards a gentle pinch configuration through quasi-static contraction of the core as a whole before the onset of the dynamical inside-out collapse of the inner core.

Various mechanisms have been proposed to remove angular momentum from the gas falling onto a protostar through a Keplerian disk within the inner core; many use the observed, ubiquitous, bipolar outflows which turn on during the Class 0 phase of star formation (Konigl \& Pudritz 2000; Shu et al. 2000; Tomisaka 1998).  The general expectation from the above magnetically regulated collapse and outflow models is that the outflow axis will be parallel or nearly so to the core magnetic field, since in these models the core rotation axis is aligned or nearly aligned with the core magnetic field, and so the Keplerian disk and bipolar outflow axes are also parallel or nearly parallel to the core magnetic field lines.    The cartoon and caption in Figure 1 summarize the above magnetically regulated scenario.

The basic scenario depicted in Figure 1 has not been observationally verified;  so far observations tailored to test parts of this scenario have not provided irrefutable evidence for the general existence of pseudo-disks, let alone the alignment of the mean magnetic fields with respect to the symmetry axis of these disks.  Recent polarimetry studies by Ward-Thompson et al.\ (2009) and Tassis et al.\ (2009) imply mean magnetic field directions more aligned with the short axis of observed flattened cores than not (consistent with Figure 1), but by no means aligned.  The statistical analysis of Tassis et al.\ (2009) of their 24 high-mass star forming regions assumed random line-of-sight angles for the cores and magnetic fields, resulting in a best-fit model consisting of a thick oblate core with a magnetic field on average having an angle of 24$^{\circ}$ to the short axis of the core.  Tassis et al.\  (2009) state that more observations are required in order to reduce the inaccuracies caused by unknown line-of-sight projection effects, but their results do provide some validity to the pseudo-disk scenario, albeit with slightly mis-aligned magnetic field configurations.

In regards to the near alignment of the outflow to the magnetic field shown in Figure 1,
M\'{e}nard \& Duch\^{e}ne (2004) used polarization measurements of background stars to study the relative orientations of the jets from Classical T Tauri stars (CTTS) embedded in the Taurus-Auriga molecular cloud and the magnetic field in their vicinity. Typically the background stars used in this study were separated by about 0.5$^{\circ}$ or more from the evolved CTTS being examined, so the magnetic field being probed in this way is not that of the field of a remnant core surrounding the CTTS, but the magnetic field of the molecular cloud in the vicinity of the CTTS.  This study concluded that the jets of embedded CTTS in the molecular cloud Taurus-Auriga are not aligned with the large-scale molecular cloud magnetic fields.   Indeed the jets within this one molecular cloud are not aligned with each other.  Large scale turbulence may play a role -  possibly producing angular momentum and magnetic structures within the cores which are not aligned with the major axis of symmetry or the large-scale magnetic field of the much larger parent molecular cloud.  Possible evidence for this can be seen in Taurus-Auriga in the two pre-stellar cores observed by Kirk, Ward-Thompson, \& Crutcher (2006), for which the polarized 850 $\mu$m emission indicates changes in magnetic field directions in the cores relative to the large-scale magnetic field in the cloud.  It is the magnetic field in these cores,
not the large-scale magnetic field,
that may align the direction of future outflows from that core if magnetic regulation occurs within the cores. 
 Hence the results of M\'{e}nard \& Duch\^{e}ne (2004) do not necessarily disprove the magnetically regulated scenario.

Vall\'{e}e, Bastien \& Greaves (2000), Henning et al.\ (2001), Matthews \& Wilson (2002), Vall\'{e}e, Greaves \& Fiege (2003),  Wolf, Launhardt \& Henning (2003), 
 Girart, Rao \& Marrone (2006), Kwon et al.\ (2006) and Attard et al.\ (2009), among others, 
 undertook observational studies of magnetic field structures within cores containing low-mass, embedded YSOs
with bipolar outflows
using polarized submillimeter  continuum emission. Low-mass YSOs are good objects to study because they tend to be embedded in simple, relatively isolated regions.  The results of the above studies taken as a whole do not show a clear case for the alignment of outflows with core magnetic fields.  However, most of the YSOs in this combined sample are binaries, are very distant, are not Class 0, or do not have their outflow axis lying close to the plane-of-the-sky.  The last criterion is an important one because a pinched magnetic field structure, which is symmetric about an axis of an outflow with a large line-of-sight component, would produce polarization vectors with a large variation in position angles.   To minimize projection effects when testing the theory of alignment between outflows and core magnetic fields, the axis of the outflow should lie close to the plane-of-the-sky.

We have begun a survey of low-mass, isolated (single), nearby ($\le$ 400 pc), Class 0 YSOs that have well defined bipolar outflows which lie nearly in the plane-of-the-sky.   Our study will provide maps of the 350 $\mu$m polarization vectors within a 10,000 AU radius around each embedded YSO in our survey. We will also provide interferometric maps of the outflows within the cores for each of our survey YSOs where such maps do not already exist. This is so we can better determine the orientation of the outflow and its interaction with the core gas.  
We are using 350 $\mu$m rather than longer wavelength polarization because we want to obtain the best spatial resolution using SHARP (see section 2), 
and because 350 $\mu$m emission is weighted towards the warmer emission of the cores immediately surrounding a YSO rather than towards the extended cooler core envelope. This paper gives a summary of our results
on three of the YSOs in our study, L1527, B335, and IC348-SMM2.   
Section 2 outlines our observations and summarizes our results, Section 3 compares these results to magnetically regulated models, and Section 4 gives our conclusions.

\section{Observations and Results}

\subsection{Polarimetry Observations and Data Analysis}
The observations presented here were carried out at the Caltech Submillimeter Observatory (CSO) using the SHARP polarimeter.  SHARP is a fore-optics module that provides polarimetric capability to SHARC-II, the CSO's 32 $\times$ 12 pixel submillimeter camera.  SHARP divides the incident submillimeter radiation into two orthogonally polarized beams that are then reimaged onto the two ends of the ``long and skinny'' SHARC-II detector.  A half-wave plate upstream of the polarization-splitting optics is rotated every few minutes during data collection, and the two orthogonally polarized 12 $\times$ 12 pixel ``sub-images'' acquired for four different half-wave plate angles are combined in software to determine the total flux as well as the linear polarization state of the radiation.  SHARC-II and SHARP are described respectively by Dowell et al.\ (2003) and Li et al.\ (2008).   The 350 $\mu$m observations described here were made with a beam size of 10$\arcsec$ (FWHM).  

The data were collected in chop-nod mode.  This involves rapidly modulating the tilt of the CSO secondary mirror in cross-elevation (``chopping'') while more slowly ``nodding'' the entire telescope back and forth, thereby making near simultaneous measurements of the source and two nearby sky reference positions.  A chop-nod observation is carried out at each of the four half-wave plate angles (0$^{\circ}$, 22.5$^{\circ}$, 45$^{\circ}$, and 67.5$^{\circ}$) thereby forming a ``half-wave plate cycle''.  As described by Dowell et al.\ (1998) and Kirby et al.\ (2005), the net effect is that the total source flux and a component of the linearly polarized flux are measured for each half-wave plate position in a cycle, while removing the spatially extended ``background'' that includes atmospheric and telescope emission as well as any Galactic emission covering a sky area large compared to the chopper throw.  For the observations described here, the chopper throw ranged from 120$\arcsec$ (for B335 and L1527) to 300$\arcsec$ (for IC348-SMM2) and the chopping frequency was $\sim$1 Hz.  Because of sky rotation, the reference beams rotate about the main beam on the sky as the source is tracked.  Our three targets do have extended flux, so it was necessary to avoid making observations at hour angles for which significant source flux contaminates the reference beams.  

Each SHARP half-wave plate cycle requires about 7 minutes of elapsed time.  In between successive half-wave plate cycles, the pointing position is dithered by about 10-20$\arcsec$.  Data were obtained for B335 during the nights of 28, 29, and 30 April 2007 ($\sim$80 cycles; zenith submillimeter opacity $\tau_{350 \mu m} \sim$ 1.0 - 1.8), for IC348-SMM2 on the night of 10 November 2007 ($\sim$20 cycles; $\tau_{350 \mu m} \sim$ 1.0), and for L1527 during the nights of 9 and 10 November 2007 and 6 and 10 September 2008 ($\sim$85 cycles, $\tau_{350 \mu m} \sim$ 0.8 - 1.2). 

SHARP data analysis is carried out in two steps.  In the first step, each individual half-wave plate cycle is processed to obtain three 12 $\times$ 12 pixel maps, one for each of the Stokes parameters $I$, $Q$, and $U$.\  (Parameter $I$ corresponds to the total flux while $Q$ and $U$ fully characterize the linearly polarized flux.)  A detailed discussion of the techniques used during this first step can be found in Section 3.4 of Hildebrand et al.\ (2000) and Equations 2, 3, and 4 of Attard et al.\ (2008).  Corresponding error-maps are also obtained for each of the three Stokes parameters.  This is done by first using the variance in the individual total and polarized flux measurements to estimate the uncertainties in this time stream (as in section 4.1 of Dowell et al.\ 1998) and then propagating these uncertainties to the Stokes parameter maps.  At this point, we remove from the Stokes parameter maps any pixels that have abnormally high errors.  

The second step of the analysis involves combining these single-cycle 12 $\times$ 12 pixel maps of $I$, $Q$, and $U$, to form final maps of $I$, $Q$, and $U$.  We account for the dithering and the sky rotation by interpolating the single-cycle maps onto a regular equatorial-coordinate grid, which causes a modest loss of angular resolution (Houde \& Vaillancourt 2007).  Errors in the final maps are propagated from the corresponding errors in the single-cycle maps.  Corrections for changing atmospheric opacity (Kirby et al.\ 2005) as well as for instrumental polarization and polarimetric efficiency (Li et al. 2008) are also made during this second analysis step. 

An important question that we can ask during the second step of the analysis is whether the single-cycle $Q$ and $U$ maps are consistent with one another within their nominal errors.  Recall that these nominal errors are computed during the first step of the analysis.  A quantitative answer to this question is provided by the reduced chi squared, $\chi_r^2$.  Averaging together the $\chi_r^2$ values found over the $Q$ and $U$ maps of each source, we obtain mean $\chi_r^2$ values of 1.7 for L1527, 1.5 for IC348-SMM2, and 2.1 for B335.  We do not know the origin of the extra errors that cause these elevated $\chi_r^2$ values.  However, for each of our three data sets (one for each source) we were able to verify that these errors occur mainly on time scales that are short compared to the total duration of the data set.  (This duration is several hours for IC348-SMM2, and several days or even longer for the other two sources.)  Thus, it seems reasonable to treat the extra errors as if they are statistical in nature.  Accordingly, we inflate our nominal errors by the square root of $\chi_r^2$.  

Finally, the degree $P$ and angle $\phi$ of polarization and their associated errors $\sigma_P$ and $\sigma_\phi$ are computed for each sky position via standard techniques (see Section 3.4 of Hildebrand et al.\ 2000).   
The uncertainties $\sigma_P$ and $\sigma_\phi$ are affected by both the polarized flux errors ($\sigma_Q$,  $\sigma_U$) and the total flux errors ($\sigma_I$).  The latter have a negligible effect, however, because $P << 100\%$.  
We consider sky positions for which $P \ge 2\sigma_P$ to be polarization detections and sky positions for which $P + 2\sigma_P < 1.0\%$ to be low upper limits on $P$.\  $P$ is then corrected downward to account for polarization bias (as described in Section 4.2 of Hildebrand et al.\ 2000 and in Vaillancourt 2006).   We keep track of the sky positions where the polarimetric significance drops below $2\sigma$ after this bias correction, as discussed below.  

Note that we have chosen to set our detection threshold at $P \ge 2\sigma_P$ rather than applying the more conservative criterion $P \ge 3\sigma_P$ used for previous SHARP observations (e.g., Attard et al.\ 2009).  When using the latter, more conservative threshold, the $1\sigma$ uncertainties $\sigma_{\phi}$ in the angles of polarization (which translate into $1\sigma$ uncertainties in the magnetic field angles) are below $\sim9.5\degr$ (Serkowski 1962).  With our more lenient threshold, these $1\sigma$ error bars range up to almost $15\degr$ (see Table 1).  Because our goal in this paper is to test the gross predictions of the magnetically regulated collapse models rather than its fine details, we believe that this degree of uncertainty is acceptable.  However, it should be kept in mind when interpreting our polarimetry results.

\subsection{Polarimetry Results}
Our polarization detections for L1527, IC348-SMM2, and B335 are presented in Table 1. 
Figures 2a, 3a, and 4a illustrate the results listed in Table 1.  Contours indicate the total intensity, I, measured by SHARP.  The highest contour level corresponds to 90\% of the peak flux, and each subsequent contour represents a decrement of 10\% of the peak flux. 
Except for very narrow strips at the map edges, all sky positions mapped have total flux errors well below 10\% of the peak flux.  The morphologies seen in our total intensity contour maps thus represent real structures, with the exception of a few very small flux peaks and flux holes seen at the map edges (e.g., southwest edge of Figure 3a, and northeast and northwest edges of Figure 4a).  All three sources have been mapped at 850 $\mu$m (Chandler \& Richer 2000, Hatchell et al.\ 2005, J{\o}rgensen et al.\ 2007), and the morphologies seen at this longer wavelength are similar to those seen in our SHARP maps. 

The bars in Figures 2a, 3a, and 4a indicate polarization detections.  The length of each bar is proportional to $P$ (see key at bottom left of each figure) and its orientation indicates the direction of the E-vector of the polarized radiation.  The thickness of each bar shows its significance: sky points having greater than  $3\sigma$ significance after bias correction are thickest, points with greater than $2\sigma$ post-correction significance but less than $3\sigma$ have intermediate thickness, and sky points for which the significance drops below $2\sigma$ after bias correction are shown using the thinnest bars.   For two sky positions, we obtained $2\sigma$ upper limits below 1.0\%.  These correspond to the peak of B335, where we find a $2\sigma$ limit of 0.6\% and a position offset from the peak of L1527 by ($\Delta \alpha, \Delta \delta$) = (+9.5$\arcsec$, -9.5$\arcsec$), where our $2\sigma$ limit is 0.9\%.
Although we do not make use of these upper limits in this paper, we include them in our figures since they represent sensitive measurements that may one day be useful.  The two limits are indicated with open circles in Figures 2a and 4a.  

Comparing our polarization results for B335 with the 850 $\mu$m polarimetry of this source presented by Wolf, Launhardt \& Henning (2003; hereafter WLH03), we find reasonable agreement.  The WLH03 map is discussed further in section 3.2.2 below.  Similarly, the 850 $\mu$m polarization map of L1527 that is presented with no interpretation in the SCUPOL archive paper (Matthews et al.\ 2009) agrees with our 350 $\mu$m polarization map of this source.  In making these comparisons, we refer only to the measured angles of polarization since the degree of polarization of thermal dust emission is known to generally have considerable wavelength dependence (Vaillancourt et al.\ 2008, and references therein).

\subsection{B335 Outflow Observations and Data Analysis}
The B335 350 $\mu$m emission is very compact (see Figure 4a).  To date, outflow maps for B335 have not been made with sufficient spatial resolution to determine the structure and kinematics of the outflow within the compact core region of B335. As part of our survey, we measured the outflow within the B335 core using the CARMA interferometer.
	
The CARMA observations of B335 were made with the 15-element array in the D-configuration
in July 2008. The total duration of the observations was 8.4 hours. The QSO 1925+211 was
the phase calibrator; its flux was determined to be 1.2 Jy, using both Uranus and Neptune
as flux calibrators. Observations of 3C454.3 were used to calibrate the passband
structure.  The correlator was configured to place the CO 1-0 line in the upper sideband; with 63
channels across a 7.6 MHz bandwidth, the velocity resolution at 115.271 GHz is
approximately 0.32 km\,s$^{-1}$. Phase calibration was carried out on observations of the 2.7mm
continuum emission in two 500 MHz bands. The gains determined for these wide (continuum)
bands were then applied to the narrow (line) bands.

The data were calibrated and mapped using the MIRIAD software package. The Fourier
transform of u,v visibilities was taken, constrained by an image cell size of 1$\arcsec$ and a
natural weighting function. The size of the synthesized beam is 4.4$\arcsec$ $\times$ 3.8$\arcsec$.
The peak intensity in the continuum emission (Figure 4b, grayscale) is 16.8 mJy/beam; the
rms noise is 1.8 mJy/beam. The blue and red contours in Figure 4b show the CO 1-0 line
emission integrated over the velocity ranges 5.4 - 8.0 km\,s$^{-1}$ and 8.6 - 11.2 km\,s$^{-1}$,
respectively. The contours are at intervals of 3$\sigma$, where $\sigma$ is 174 mJy/beam $\times$ km\,s$^{-1}$. The integrated intensity peak in the blueshifted outflow is 5.6 Jy/beam $\times$ km\,s$^{-1}$; the peak in the redshifted outflow is 4.0 Jy/beam $\times$ km\,s$^{-1}$. 

\section{Discussion}

\subsection{Core Orientation with respect to the Outflows}

Magnetically regulated models predict that a YSO will be surrounded by a pseudo-disk a few thousand AU in size, and that the pseudo-disk symmetry axis will be aligned or nearly so with that of the YSO bipolar outflow. In this sub-section we will explore the extent to which this is the case for our three sources.

In the case of L1527, Hogerheijde \& Sandell (2000) used a two Gaussian model to separate the 450 and 850 $\mu$m emission of an inner compact core immediately surrounding the embedded YSO from the more extended emission.  Their Table 3 shows a deconvolved FWHM compact core size of 10$\arcsec$ $\times$ 5$\arcsec$ (1400 AU $\times$ 700 AU) with a position angle of the major axis $\sim$ 30$^{\circ}$ east of north.  However, the interferometric C$^{18}$O map of this core by Ohashi et al. (1997; see Figure 5a), made with slightly higher spatial resolution 
and with the advantage that the extended emission is automatically removed, implies the major axis of the core is oriented north-south, but that the outer regions of the flattened core are distorted -- possibly by the bipolar outflow -- thus giving an impression of a tilt away from the north-south orientation when mapped with slightly poorer resolution and extended emission confusion. The bipolar outflow measured by Hogerheijde et al. (1998) in $^{12}$CO and by Zhou, Evans \& Wang (1996) in $^{13}$CO has a position angle of 90$^{\circ}$ east of north.  Thus L1527 is observed to have a flattened core on the scale of a few thousand AU, consistent with an edge-on pseudo-disk with its symmetry axis aligned with the outflow axis.

In the case of B335, Chandler  \& Sargent (1993) observed a highly flattened core at 2.7 mm that is resolved only along the major axis, which has a size of $8\farcs2$ (2000 AU) with a position angle of 20$^{\circ}$ east of north (see Figure 7a).  The bipolar outflow measured in $^{12}$CO by ourselves in this paper, Hirano et al. (1988) and others has a position angle of 90$^{\circ}$ east of north.   Thus B335 is observed to have a flattened core on the scale of a few thousand AU, consistent with an edge-on pseudo-disk with its symmetry axis slightly tilted relative to the outflow axis by about 20$^{\circ}$.

The core emission from IC 348-SMM2 has not previously been observed in detail.  We use our data to study the morphology of the core surrounding this source.  Figure 3a shows the underlying 350 $\mu$m continuum emission intensity from IC348-SMM2 which we measured with SHARP. The 50\% 350 $\mu$m contour of this source has a major axis with a position angle of 60$^{\circ}$ east of north.  The diameter of this major axis is 27$\arcsec$, compared to the perpendicular minor axis diameter of 20$\arcsec$.  However, the 50\% contour probably does not represent the FWHM of the core surrounding IC348-SMM2.  A more realistic estimation of the FWHM of the core would be the 65\% contour, which represents the 50\% contour level of the core after the subtraction of a more extended cloud background set at about the 30\% contour level in Figure 3a. The 65\% contour major axis has a position angle of 56$^{\circ}$ east of north, and the diameters of the major and minor 65\% contours are 21$\arcsec$ and 15$\arcsec$, respectively. If we assume an elliptical Gaussian core, we can deconvolve these FWHM estimates with a 10$\arcsec$ FWHM Gaussian beam to get a core size of 18$\arcsec$ $\times$ 11$\arcsec$  (5500 AU $\times$ 3300 AU) at a position angle of 56$^{\circ}$. The size measured here is larger than for a typical pseudo-disk quoted in the literature by about a factor of two, but at our spatial resolution we are probably not resolving the true FWHM of the pseudo-disk, we are more likely measuring the outer extended region.  The bipolar outflow measured by Tafalla, Kumar \& Bachiller (2006) in $^{12}$CO has a position angle of 17$^{\circ}$ west of north.  Thus IC348-SMM2 is observed to have a flattened core on the scale of several thousand AU, consistent with the outer regions of an edge-on pseudo-disk with its symmetry axis tilted within 20$^{\circ}$ relative to the outflow.

In summary, we do see evidence for pseudo-disks and a tendency for alignment, within 20$^{\circ}$, of the pseudo-disk symmetry axis and the outflow axis for each of the YSOs in our sample.

\subsection{Inferred Magnetic Field Structures}

Our submillimeter polarization measurements displayed in Figures 2a, 3a, and 4a do not give a measure of the strength of the magnetic field, but they do give an indication of the net magnetic field direction (Lazarian 2007) and the level of uniformity (Hildebrand et al. 2009) along a given line-of-sight.   The net magnetic field direction along a given line-of-sight is perpendicular to the submillimeter polarization vector measured for that line-of-sight (Lazarian 2007).   
Figures 2b, 3b, and 4b display the direction of the magnetic fields  surrounding L1527, IC348-SMM2 and B335, respectively, as determined by the polarimetry $E$-vectors rotated by 90 degrees.  
We compare these field directions with our observations of the cores' density structure (i.e., pseudo-disk) and velocity structure (i.e., outflows, infall radii). We note, however, that our polarization data for the three cores likely contain contributions from both the magnetic field associated with the core under study as well as that associated with the larger surrounding cloud.  To see this, recall (section 1) that the maps shown by Kirk et al.\ (2006) reveal large changes in field direction as one moves away from the center of a core into the surrounding cloud material; and note that these changes occur at positions located about one arcminute ($\sim$8000 AU) from the centers of the Kirk et al.\ (2006) cores, while our own maps extend out to distances of 7000-15,000 AU from the central YSO.  Thus a significant fraction of the emission we have observed may originate from the cloud at large, not the core under study.

One way to account for such contamination is to flag polarization measurements made at positions having lower flux density.  Such measurements could be contaminated to a large degree by line-of-sight emission associated with the larger parent cloud.  The fluxes we measure at the edges of our three maps range from $\sim$5\% to $\sim$40\% of the respective peak flux (see contours in Figures 2a, 3a, and 4a).  We have chosen to set a ``contamination threshold'' at 25\% of peak flux; we will consider polarization measurements obtained at positions having flux below this threshold to be at risk of large amounts of contamination by polarized signals originating in the larger cloud.  This choice is somewhat arbitrary, but it has the benefit of flagging polarization measurements coming from regions of very low flux while preserving most of our measurements; we flag as unreliable just four measurements out of 22.   All of our $B$-vectors shown in Figures 2b, 3b, and 4b are drawn with the same length and thickness, and those associated with total intensity levels below our contamination threshold are shown using dashed lines.  Figures 2b and 4b each contain one such suspect $B$-vector, and Figure 3b contains two.

Also shown in Figures 2b, 3b, and 4b are the bipolar outflow morphologies in these sources (see the figure captions for references).   In addition, these figures show circles depicting the measured outer limits of the infall regions in L1527 and B335 based on inverse P-Cygni line profiles (see captions for references).  
No measurement of the infall radius for IC348-SMM2 has been made to date, but such an infall region is likely to have a radius range of 20$\arcsec$ to 30$\arcsec$ for an infall age similar to the ages of L1527 and B335 (see Table 2).  Below we will assume an infall radius of 25$\arcsec$ when we compare our observational results to models.

The models we will compare our data to are those of ASL03a and ALS03b.  These models consist of self-similar, self-gravitating, singular, isothermal toroids with various amounts of rotation and magnetization.  The rotation speeds of the cores range from 0 to 0.5 times their thermal sound speeds and the magnetic-flux-to-mass-ratios of the cores range from from 0 to 0.5.  All models are supercritical in order for collapse to occur without external pressure. Presumably the collapse phase occurs after ambipolar diffusion has occurred, producing the supercritical state in the core. ALS03b note, however, that even in the relatively weakened state of the fields, these fields are responsible for the formation of pseudo-disks, considerable transport of angular momentum, and the resulting size of the centrifugally supported Keplerian disk during the collapse phase, and so cannot be ignored.  In Figures 5, 6, and 7 we compare our data on L1527, IC348-SMM2, and B335, respectively, to the model displayed in Figure 8c of ALS03b, a model with intermediate rotation and magnetic field strengths. It should be noted here that, except for the case where there is no magnetic field to flatten the core and provide polarization, the spatial resolution of our data precludes us from discerning between the various models presented in ASL03a and ALS03b (see Figures 7 and 8 in ALS03b). ALS03b show (their Figure 7) that rotation has only a minor effect on the gas and magnetic field geometry at the spatial scales we are measuring here. However, the aim of our current study is not to test the finer points of magnetic regulated collapse models, but the gross predictions represented in Figure 1 and evidence for magnetic field pinches.

In addition to our spatial resolution constraints, it is important to bear in mind that our results represent an integration of polarizations along each line of sight, whereas the magnetic field given in Figure 8c of ALS03b represents only the cross-section of the poloidal field on the plane of the sky for an edge-on pseudo-disk.  If this cross-section were rotated round the symmetry axis of the pseudo-disk, an integration along a line-of-sight would result in a weakening of the pinch geometry to a more uniform field aligned with the symmetry axis. 

\subsubsection{Magnetic Field Structure around L1527 and IC348-SMM2}

Scrutiny of the results displayed in Figures 2 \& 3 reveals that the field structures in L1527 and IC348-SMM2 are generally consistent with the magnetically regulated dynamical collapse models cited in the introduction in that they show: (1) pinched field line vectors on the scale of the measured or inferred infall regions for these cores; and (2) field line vectors (with a few exceptions discussed below) that are basically aligned with the bipolar outflows (once the distortion of a pinch is subtracted by eye using Figures 5 \& 6). The exceptions mentioned in (2) are the three polarization vectors in the low-flux region to the south of IC348-SMM2, which imply an east-west magnetic field, and a vector immediately east of the emission peak of L1527. A possible explanation for this latter vector is given later in this section. However, the east-west field lines in IC348-SMM2 cannot be explained in the context of a magnetically regulated model.  It is possible these vectors may not be associated with the core of IC348-SMM2, since two of the three reside in a region which has emission less than 25\% of the peak emission for the source, and the third resides in a region with emission that is 25.5\% of the peak.  

Although in general the polarization results agree with magnetically regulated models, in detail we see significant discrepancies beyond the exceptions mentioned above. The scale of the pinched structure in magnetically regulated models depends on the size (i.e., age) of the infall region.  Outside the infall region, the magnetic fields should be uniform or nearly uniform, depending on the details of the pre-collapse, quasi-static contraction of the core. Only within the infall regions should pinches of the field lines be significant.  

Figure 6 shows how our $B$-vectors in IC348-SMM2 (with our assumed infall radius and ignoring the three $B$-vector exceptions to the south) show a remarkable agreement with Figure 8c from ALS03b if the symmetry axis of the pseudo-disk has a 17$^\circ$ tilt with respect to the outflow as we measure for this source. The agreement of the predicted field geometry from the ALS03b model with our measured vector position angles is somewhat surprising since ALS03b Figure 8c gives a cross-section of the poloidal field on the plane of the sky for an edge-on pseudo-disk, while our measurements are an integration along the line of sight. Such an integration should smooth out the pinched structure to some degree.  The fit by eye between the model and our measurements is not as good if the pseudo-disk symmetry axis in the model is instead aligned with the outflow axis. The fact that the magnetic field fit is better when the model is aligned to the measured pseudo-disk axis, rather than the outflow axis, is consistent with the hypothesis that the pinched magnetic field structure and the pseudo-disk are both products of magnetized contraction and collapse, whereas the outflow is probably a product of rotation.  In IC348-SMM2, the rotation axis may not be exactly aligned with the overall magnetic field axis for the core.

Figure 5 shows that in L1527 our $B$-vectors broadly agree with a pinched magnetic field structure aligned with the pseudo-disk axis as shown in Figure 8c of ALS03b, but the figure also shows that there are significant distortions beyond the uncertainties ($\pm$11$^{\circ}$) in the angles of the vectors.   We see these distortions close to the edge and in one case beyond the infall boundary.  A possible explanation for this additional distortion could be the bipolar outflow in this source which overlaps a significant portion of the region containing our measured vectors (see Figure 2). The effects of bipolar outflows are not included in the models to which we are comparing our results (ASL03a; ALS03b; Galli \& Shu 1993a,b).  The bipolar outflow in L1527 may also be responsible for the exceptional vector identified previously, which lies immediately east of the L1527 emission peak. This vector implies a magnetic field direction that is almost perpendicular to the outflow. In this scenario, the outflow pushes core material, and therefore also core magnetic field lines, into two polar cones surrounding the bipolar outflow near the emission peak in L1527, thereby giving rise to both the additional distortions near the edge of the infall boundary as well as the exceptional vector just east of the peak.  Evidence for conical cavities such as would be required in this scenario has been obtained via mid-IR scattered light observations by Tobin et al.\ (2008) as well as interferometric measurements of HCO$^+$ by Hogerheijde et al.\ (1998).  The latter is shown in Figure 2b.  In addition, the submillimeter maps of Hogerheijde \& Sandell (2000) and our 350 $\mu$m map (Figure 2a) show evidence of an X-like structure in the extended background emission about L1527 that overlaps the observed outflow in that region, implying that significant submillimeter emission (and any polarization of such emission) must be coming from the surface of the outflow conical cavities.

Could the bipolar outflow observed in L1527 have enough energy to distort the magnetic field structure linked to the gas it entrains? The diagrams in Figure 8 in ALS03b show the pinched magnetic field structures for a number of rotating, dynamically-collapsing toroids of different initial cloud magnetic field strengths.  These diagrams also show the contours of $\beta$ (the ratio of the thermal to magnetic energy density).   Close to the axis of symmetry $\beta$ is much less than one ($\sim$ 0.1), further from the axis of symmetry $\beta$ increases to values above one.  These values of $\beta$ can be used together with estimates of the thermal energy within a core to determine the energy density required of an outflow to distort the magnetic field within that core.  The core thermal energy density 
for L1527 can be approximated by ${3 \over 2} \rho c_s^2$, where $\rho$ is given by the mass density of the 10$^6$ cm$^{-3}$ molecular gas measured for the L1527 core about 10$\arcsec$ away from the center (Hogerheijde \& Sandell 2000) and c$_s$ is the sound speed which is approximately 0.25\,km\,s$^{-1}$ (Zhou, Evans \& Wang 1996).  The lower limit to the outflow energy density in L1527 is expressed as ${ 1 \over 2} M_{_L} V^2_{_L}$ divided by the observed volume of the outflow lobes, where the measured values for $M_{_L}$ (the mass entrained in the outflows) and $V_{_L}$ (the velocity of the entrained gas in the outflows) are given in Table 2 from Hogerheijde et al. (1998).  The volume of both outflow lobes can be approximated by ${2 \over 3} \Omega_{_L} R^3_{_L}$, where $R_{_L}$, the extent of one lobe in L1527, is given in Table 2 and $\Omega_{_L} \approx$  0.14 sr based on Figure 6 in Hogerheijde et al. (1998) showing the extent of the $^{12}$CO outflow in L1527. 
The above imply that the energy density in the outflow is greater than the thermal energy density in the region 10$\arcsec$ from the peak of L1527 by about a factor of 100.   Since in the ALS03b models the magnetic energy density is less than the thermal energy density away from the axis of symmetry, this implies that the outflow does have the energy required to distort the magnetic field where it disturbs the gas in the core away from the axis of symmetry.  Indeed, with the energy density of the outflow observed, even the magnetic field very close to the axis of symmetry could be distorted.   

The ratio of the outflow energy density to core thermal energy density in IC348-SMM2 is similar to the ratio in L1527, if one assumes similar gas densities and sound speeds for IC348-SMM2 as measured for L1527 and the outflow parameters in Table 2 for IC348-SMM2.   But the outflow in IC348-SMM2 does not overlap with our measured polarization vectors to a very great degree, so the outflow cannot affect the alignment of the magnetic fields which are inferred by these polarization measurements.   Where there is overlap, the distortion would be minimal for field lines already parallel to the axis of the outflow if the opening angle of the outflow at that location is small -- as it is for IC348-SMM2 ($\pm$15$^{\circ}$; see Figure 3). 

In summary, if one assumes that the low-flux exceptions in IC348-SMM2 are not part of the core and one includes the effects of bipolar outflows on field alignment, then our observations of L1527 and IC348-SMM2 are consistent with magnetically regulated models.

\subsubsection{Magnetic Field Structure around B335}

At first glance,  our two polarization vectors in B335 imply a magnetic field more perpendicular than parallel to the outflow axis.  However,  our two inferred $B$-vectors are not too different from what would be expected in the region just south-east of the B335 center based on Figure 8c of ALS03b when this figure is aligned with the minor axis of the flattened core as measured by Chandler \& Sargent (1993) (i.e., the symmetry axis of the inferred pseudo-disk).  In this region of the model, the field lines are pinched towards the axis of symmetry (see Figure 7).  

Figure 7 also shows the $B$-vectors inferred from the 850 $\mu$m polarimetry measurements of WLH03 as well as those inferred from our 350 $\mu$m polarimetry measurements.  Our results are broadly consistent with the results of WLH03, in that they agree with four of the six vectors in the south-east quadrant region of the B335 core, but not, however, with the one vector with which our measurements most nearly overlap.  The 20 $B$-vectors of WLH03 and our 2 all lie inside the infall region outlined in Figures 4 and 7.  The WLH03 results imply on average a more N-S magnetic field structure within B335, but there is considerable distortion evident in the field lines implied by the 20 vectors,
since the standard deviation of  the position angles of the $B$-vectors is about three times the average measurement error for each vector.
 WLH03 concluded that the average field they measure in B335 is the direction of the field in the core when it collapsed, and that the flattened core seen in B335 is prolate (rather than oblate) with its symmetry axis parallel to the magnetic field. 
If this is the case, then B335 presents a counter example to the results obtained by Tassis et al.\ (2009)  who concluded the core model that best fits their sample of 24 high-mass cloud cores is an oblate core with the mean magnetic field more aligned with the short axis than the long axis.  

Alternatively, the WLH03 data and ours could imply a more toroidal field in B335 than poloidal.  But if this were the case then this is inconsistent with the model of ALS03b which includes rotation.  In this model, only a small volume of a dynamically collapsing core contains twisted (i.e., toroidal) magnetic fields.  Outside this small region the collapsing flow has yet to be spun up, and inside this region $\beta \ll 1$ so the field lines are rigid.  If B335 does contain a toroidal field configuration, then the ALS03b model fails to describe it; there appears to be more twisting of the original poloidal magnetic field lines than can be explained in their model.  

Yet another interpretation of the data, and the one we advocate in this paper, is that B335 is an extreme example of what is happening in L1527; the outflow in B335 has distorted the field lines in the core of B335, either directly or indirectly by exciting more turbulence within the core, so the magnetic field seems to align N-S on average.  Could the outflow in B335 cause large field distortions in the core? If we use the same analysis we used for L1527, but with a core thermal speed of 0.23\,km\,s$^{-1}$ (Zhou et al.1993), a core density at about 10$\arcsec$ from the center of $\sim$ 10$^5$ cm$^{-3}$ ( Zhou et al. 1990; Harvey et al. 2003), the B335 values for the outflow parameters given in Table 2, and $\Omega_{_L} =$ 0.6 sr (Figure 2 in Hirano et al. 1988), then we find that the outflow energy density is about a factor of 6 higher than the core thermal energy density (10$\arcsec$ from the peak) in B335. Our own outflow data shown in Figure 4b give a total kinetic energy for the outflow within the region mapped (i.e., ${1 \over 2}M_{_{chan}} V^2_{_{chan}}$, summed over the velocity channels in the ranges 5.4 - 8.0 km\,s$^{-1}$ and 8.6 - 11.2 km\,s$^{-1}$) of 1.8 $\times$ 10$^{42}$ erg, once a correction for the 10$^{\circ}$ inclination of the outflow to the plane of the sky has been made. Dividing this by the volume of the outflow as defined in Figure 4b, we get a kinetic energy density of $\sim$ 10$^{-9}$ erg cm$^{-3}$, which is a factor of 3 higher than the core thermal energy density. This outflow energy density although not enough to distort the magnetic field close to the axis of symmetry of the pseudo-disk in B335, is enough to distort the field further out. 

The outflow cavity outlined by the CO observations in Figure 4b is coincident with the region within the B335 core that most likely would have a magnetic field energy density that is greater than the outflow energy density; beyond this outflow cavity the reverse is likely to be true.  The outflow cavity is a region in which the gas and dust densities are low, and so it is a region of low 350 $\mu$m emission along our line of sight.  Our polarization measurements are thus weighted away from the outflow cavity region towards those regions along our line of sight where the energy density of the outflow could distort the magnetic field structure.

But why is this distortion observed to be so large in B335 compared to the distortion in L1527?   The difference between  B335 and the other two YSOs presented here is that the bipolar outflow in B335 is much larger in length, width and apparent age than the outflows in L1527 and IC348-SMM2 (Table 2).  Therefore, the field lines within the B335 core could be highly distorted because the outflow has had time to plow through a greater portion of the core or excite greater gas turbulence in the core.  

\subsubsection{Chi-Square Tests of Various Magnetic Field Geometries}

In order to give some quantitative assessment of the magnetic field scenarios discussed above, and how they compare to other possible configurations, we carried out reduced chi-squared tests of our data using a number of different theoretical magnetic field configurations for each source.  For each source we compared the $B$-vectors implied by our data to: (1) a uniform magnetic field model where the angle of the field is aligned with the mean $B$-vector angle implied by our data for that source; (2) a uniform magnetic field model where the angle of the field is aligned with the outflow; (3) a uniform magnetic field model where the angle of the field is aligned with the symmetry axis of the observed pseudo-disk for that source; (4) a uniform magnetic field model where the angle of the field is aligned with the major axis of the observed flattened core (or pseudo-disk) for that source; and (5) a pinched magnetic field aligned with the symmetry axis of the observed pseudo-disk as presented in Figure 8c of ALS03b.  Note that model (4) also corresponds to the case where the magnetic field is toroidal.  Table 3 summarizes our results.  Each number in the table represents the reduced chi-squared value, $\chi^2_r$, for the data specified for that column against the model for that row.  The reduced chi-squared is calculated as $\chi^2_r = {1 \over {\nu}} \Sigma_i ((\theta_i - \theta_{Mi})^2 / \sigma_i^2)$ where $\theta_i$ are the data representing the angles of the $B$-vector at various locations, i;  $\sigma_i$ is the uncertainty in each data angle; $\theta_{Mi}$ is the angle of the magnetic field at the location of each data point for a particular model; and $\nu$ is the number of degrees of freedom for the data set. For these calculations, the values of $|(\theta_i - \theta_{Mi})|$ were constrained to be $\le$ 90$\arcdeg$ since our $B$-vectors have been derived from our polarization $E$-vectors which are invariant under 180 degree rotations.  For cases where $|(\theta_i - \theta_{Mi})| > 90 \arcdeg$, a value of $|[(\theta_i - \theta_{Mi} \pm 180)]| \le 90\arcdeg$ was substituted.   For B335, we carried out the same calculations using the data of WLH03 based on their Figure 1.  

$\chi^2_r $ should be close to 1 for a good fit.  An inspection of Table 3 shows that $\chi^2_r$ is close to this value only for the ALS03b model for IC348-SMM2 and B335 when we consider only our data minus the exceptions discussed in Section 3.2.1.  However the fits are not good for this model if we include the exceptions for IC348-SMM2, and if we include the WLH03 data for B335.  This is in agreement with our qualitative assessment above.  For L1527, although no model gives a good fit to our data, the two most favored are the ALS03b model and the uniform field aligned with the mean $B$-vector of our data for L1527 (i.e., 60$\arcdeg$ east of north). Our data do not differentiate between these two models, since our vectors lie mostly in two diagonal quadrants.  If we had more data in the other two quadrants, a $\chi^2_r $ test would be able to differentiate between these two models.  As it is, the $\chi^2_r $ results re-enforce our qualitative assessment above that although our data agree with the ALS03b model in a broad sense, there are enough deviations to it to require an explanation; possibly a disturbance to the magnetic field structure due to the outflow in L1527.

Further, based on Table 3, our data for L1527, IC348-SMM2, and B335 clearly favor the ALS03b model over the ``toroidal" models, and the ALS03b model is slightly more favored than the models with a uniform field aligned with the outflow and a uniform field aligned with the pseudo-disk symmetry axis.  This implies there is some evidence for a pinch in the magnetic field configuration for these sources.  However, the data of WLH03 for B335, in contradiction, favor the model with a uniform field aligned with the major axis of the flattened core over all the other models, although the fit is not good for this model, thus implying some other underlying magnetic field configuration must be present - possibly a configuration disrupted by the outflow in B335.  

In summary, the chi-squared results in Table 3 support the qualitative discussions we presented in sections 3.2.1 and 3.2.2.

We also compared our data to a random magnetic field model for each source.  We did this by calculating the root-mean-square (RMS) of the differences between the directions of the $B$-vectors at points lying spatially adjacent to one another in each of our sources.  For L1527 we obtained 11 pairs of measurements separated by 10$\arcsec$ or $\sqrt{2} \times 10 \arcsec$ (see Figure 2) and for IC348-SMM2 we obtained 10 such pairs (see Figure 3) for our RMS calculations.  Points separated by $\sim 10\arcsec$ (i.e., one beam diameter) or greater represent independent measurements.  For L1527 we calculated an RMS value for these differences in adjacent $B$-vector directions to be 19$\arcdeg$, and for IC348-SMM2 a value of 15$\arcdeg$.  We then carried out the same calculation on 10 randomly generated numbers ranging between -90 and 90, and did this 50 times.  The resulting RMS values formed a peaked gaussian distribution with a mean of 51 and a dispersion $\sigma = 6$.  (This mean is close to the theoretical RMS of $\sim$ 52 for randomly selected numbers between -90 and 90.) Thus our RMS values for L1527 and IC348-SMM2 lie 5.5$\sigma$ and 6$\sigma$, respectively, below the expected RMS for randomly distributed $B$-vectors, implying that the magnetic fields in L1527 and in IC348-SMM2 are not random in nature.  We carried out the same RMS calculation on two different sets of 13 pairs of adjacent $B$-vector measurements made by WLH03 in B335 (see Figure 7), resulting in RMS values of 42$\arcdeg$ and 34$\arcdeg$. Both values are much larger than the average uncertainties ($\le 10\arcdeg$) for the measurements reported in WLH03.  These RMS values imply that the magnetic field is possibly more randomized in B335 than in the other sources.  This latter point is consistent with our suggestion that the magnetic field in B335 is more distorted by its outflow than are the magnetic fields in L1527 and IC348-SMM2. 


\section{Conclusions}

We have used the SHARP polarimeter on the CSO to obtain 350 $\mu$m intensity maps and polarization measurements 
on L1527, IC348-SMM2, and B335
to test
whether or not
 magnetically regulated models for low-mass star formation 
are consistent with observations of sources which should have little distortions in their various geometries due to projection effects (i.e., sources with bipolar outflows which lie close to the plane-of-the-sky).

(1) Our data for IC348-SMM2 combined with the data of others for L1527 and B335 show flattened cores consistent with edge-on pseudo-disks having symmetry axes that are nearly, but in two cases not exactly, parallel to the bipolar outflows in these sources.

(2) There is evidence that the sources L1527 and IC348-SMM2 each contains a pinched magnetic field structure with its symmetry axis approximately aligned with the symmetry axis of the inferred embedded pseudo-disk in each. 
The evidence is strong (i.e., the goodness-of-fit to the data is good) for IC348-SMM2, if certain low-flux polarimetry measurements, which could be associated more with the background cloud than the core, are ignored.
 In IC348-SMM2, where the outflow and pseudo-disk axes are not exactly aligned, the pinched magnetic field structure  fits the data better when it is aligned with the symmetry axis of the pseudo-disk rather than with the outflow axis.  This is consistent with the hypothesis that the pinched magnetic field structure and the pseudo-disk are both products of magnetized contraction and collapse, whereas the outflow is only indirectly related to the core magnetic field structure inasmuch as that magnetic field structure has influenced the orientation of the rotation axis of the core.

(3) In L1527, however, the magnetic field structure shows considerable distortion from an ideal pinched field line structure given the measured infall radius of this source. Our hypothesis is that this distortion is caused by the bipolar outflow in L1527. We show that the outflow has sufficient energy density to distort the magnetic field structure in the core of this source.  This distortion is not seen in IC348-SMM2, because the only inferred $B$-vectors that overlap with the outflow in this source are vectors that would not have been altered by the outflow.

(4) The magnetic field structure observed in the B335 core is not aligned with the outflow axis of this source, but our two $B$-vectors are consistent with models with pinched field lines through a pseudo-disk, if the pseudo-disk is tilted with respect to the outflow axis by 20$\arcdeg$ (i.e., consistent with the pseudo-disk observed for B335 by Chandler \& Sargent (1993)).  However, if we combine our data with those of WLH03, the fit to these pinched field line models is not good.  Our explanation is that B335 is an extreme example of the bipolar outflow driven field distortion that we are seeing in L1527.  The main difference between L1527 and B335 is that the outflow in B335 is much larger in extent than the one in L1527, is assumed therefore to be much older, and so has had time to cause a greater degree of distortion of the core magnetic field lines. 
This is not the interpretation given by WLH03 of their B335 observations.  They interpreted their results to imply a near uniform field lying perpendicular to the outflow in this source - in contradiction to the predictions of magnetically regulated collapse as summarized in Figure 1.  
 
(5) More core magnetic field structures need to be mapped to elucidate the overall dynamical collapse story, given the considerable variation in core structures observed in our sample of just three.  
In short, the gross predictions of the magnetically regulated models (i.e., as summarized in Figure 1) need to be tested further.

\section*{Acknowledgements}

We are grateful to the Caltech Submillimeter Observatory TAC, management, and staff for making this study possible over the many observing runs that it has consumed.  We thank Darren Dowell, Megan Krejny, Woojin Kwon, and Hiroko Shinnaga for help with the observations, and Ron Taam for valuable discussions.  We are grateful to the National Science Foundation for supporting the Caltech Submillimeter Observatory via grant AST-0838261, and for supporting the operation of SHARP via grant AST-0909030 to Northwestern University.  Astronomical Research by P.F.G. and N.L.C. is supported by the Jet Propulsion Laboratory, California Institute of Technology.


\clearpage

\begin{deluxetable}{lrrrrrr}
\tablewidth{0pt}
\tablecaption{SHARP 350\,$\mu$m Polarimetry Results\label{tbl-specs}}
\tablehead{\colhead{Source} & \colhead{$\Delta \alpha$\tablenotemark{a}} & \colhead {$\Delta \delta$\tablenotemark{a}} & \colhead{$P$} & \colhead{$\sigma_p$} & \colhead{$\phi$\tablenotemark{b}} & \colhead{$\sigma_{\phi}$} \\
  & \colhead{(arcsec)} & \colhead{(arcsec)}   &  \colhead{(\%)} & \colhead{(\%)} & \colhead{(deg)} & \colhead{(deg)} }
\startdata
L1527 & 38.0 & 9.5  & 5.6  & 2.1 & -51.0 & 9.5 \\
& 28.5 & 19.0 & 3.3 & 1.6 & -56.1 & 12.0\\
& 19.0 & 9.5 & 2.5 & 0.9 & -63.7 & 9.7\\
& 19.0 & 19.0 & 3.5 & 1.1 & -41.7& 10.1\\
& 9.5 & -28.5 & 5.8 & 1.8 & 18.2 & 9.0 \\
& 9.5 & 0.0 & 1.0 & 0.5 & -66.4 & 12.3\\
& 9.5 & 9.5 & 1.6 & 0.5 & -36.5 & 9.6\\
& -9.5 & -19.0 & 2.8 & 1.3 & 3.1& 10.0\\
& -9.5 & -9.5 & 1.8 & 0.7 & -13.2 & 10.3\\
& -19.0 & -9.5 & 2.2 & 1.2 & -28.4 & 14.5\\
\hline
IC348-SMM2&9.5 & 19.0& 5.7& 2.1& 84.8&10.5\\
& 0.0& 19.0& 6.3& 2.4& 81.6&9.4\\
& -9.5& -28.5& 9.2& 3.7& -9.8&10.0\\
& -9.5& 19.0& 5.1& 2.8& 61.9&12.8\\
& -19.0& -28.5& 9.5& 4.4& -4.7&11.0\\
& -19.0& -19.0& 5.8& 3.0& 4.0&11.8\\
& -19.0& 19.0& 7.9& 3.4& 54.4&13.3\\
& -28.5& -9.5& 5.9& 2.9& 33.7&11.4\\
& -28.5& 0.0& 9.4& 3.1& 32.3&8.2\\
& -28.5& 9.5 & 8.0& 4.1& 26.3&12.0\\
\hline
B335 & 9.5& -19.0& 3.2& 1.5& 56.5&11.9\\
& 9.5& -9.5& 1.3& 0.7&58.8&14.6\\
\enddata
\tablenotetext{a}{Offsets from YSO positions given in Table 2}
\tablenotetext{b}{Position Angle of the polarization E-vector, measured east of north}
\end{deluxetable}

\begin{deluxetable}{lrrr}
\tablewidth{0pt}
\tablecaption{Source Information \label{tbl-specs}}
\tablehead{\colhead{} & \colhead{L1527\tablenotemark{a}} & \colhead{IC348-SMM2\tablenotemark{b}} & \colhead{B335\tablenotemark{c}} }
\startdata
RA (2000) & 04:39:53.9  & 03:43:56.9 & 19:37:01.03  \\
Dec (2000) & 26:03:11 &  32:03:06  & 07:34:11\\
\hline
 Distance (pc) & 140 & 300 & 250  \\
\hline
Infall Radius, r$_\mathrm{inf}$ (pc) &  0.026 (38\arcsec) &  \nodata  & 0.04 (33\arcsec) \\
\hline
Sound Speed, c$_s$ (km\,s$^{-1}$) & 0.25 &  \nodata  & 0.23  \\
\hline
Infall Age ($\sim$ r$_\mathrm{inf}$/c$_s$)  (yr) & $\thicksim 1 \times 10^5 $ & \nodata  & $\thicksim 1.5 \times 10^5 $ \\
\hline
$^{12}$CO Outflow Extent, R$_L$ (pc) & 0.095 (140\arcsec) & 0.14 (95\arcsec) & 0.3 (250\arcsec)  \\
\hline
PA of outflow axis E from N & 90$^\circ$ & $-17^\circ$& 90$^\circ$ \\
\hline
Outflow Mass, M$_L$ (M$_\sun$)\tablenotemark{d} & 0.18 & 0.033 & 0.44 \\
\hline
Characistic Outflow Velocity, &  &  &  \\ V$_L$ (km\,s$^{-1}$)\tablenotemark{e} & 20 & 54 & 13 \\
\hline
Outflow Age  ($ \sim$ R$_L$/V$_L$) (yr) & $\thicksim 5 \times 10^3 $  & $\thicksim 2 \times 10^3 $ & $ \thicksim 2 \times 10^4 $  \\
\enddata
\tablenotetext{a}{Zhou, Evans \& Wang (1996) for infall and Hogerheijde et al.\ (1998) for outflow information}
\tablenotetext{b}{Tafalla et al.\ (2006) for outflow information}  
\tablenotetext{c}{Zhou et al.\ (1993) for infall information and Hirano et al.\ (1988) for outflow information}
\tablenotetext{d}{Outflow Mass for both lobes combined.}
\tablenotetext{e}{Mass-weighted outflow velocities corrected for low inclination with respect to the plane-of-the-sky; $7^\circ$ (L1527), $10^\circ$ (IC348-SMM2), $10^\circ$ (B335)} 
\end{deluxetable}

\begin{deluxetable}{lcccccc}
\tabletypesize{\small}
\tablewidth{0pt}
\tablecaption{Reduced $\chi^2$ Values for Various Models\tablenotemark{a} \label{tbl-specs}}
\tablehead{\colhead{Models for $\theta_B$} &\multicolumn{2}{c} {L1527}&\multicolumn{2}{c} {IC348-SMM2} & \multicolumn{2}{c} {B335} \\
\colhead{($\theta_{Mi}$)} & \colhead{All Vectors\tablenotemark{b}} & \colhead{m/Except\tablenotemark{b}} & \colhead{All Vectors} & \colhead{m/Except} &\colhead{All Vectors} & \colhead{WLH03\tablenotemark{b}} \\ 
 \colhead{} & \colhead{($\nu$ = 10)} & \colhead{($\nu$ = 9)} & \colhead{($\nu$ = 10) } & \colhead{($\nu$ = 7)} &\colhead{($\nu$ = 2)} & \colhead{($\nu$ = 20)}}
\startdata
$\theta_{Mi}$ = Mean $\theta_i$\tablenotemark{c} & 8.3  &8.2  & 10.4   & 5.5  & 0.02  & 13.7   \\
\hline
$\theta_{Mi}$= $\theta_{outflow}$\tablenotemark{d} &  16.4  & 15.0   & 20.9  & 8.1   & 19.4   & 36.9    \\
\hline
$\theta_{Mi}$= $\theta_{PDSA}$\tablenotemark{d}& 16.4  & 15.0 & 12.7 & 4.8  &  8.2 & 38.5  \\
\hline
$\theta_{Mi}$= $\theta_{PDMA}$\tablenotemark{d}& 30.3 & 33.2 & 31.7 & 42.0 &16.2  & 14.2   \\
\hline
ALS03b Model\tablenotemark{d} & 9.9 & 7.8 & 14.0 & 0.4 & 0.8 & 31.5   \\
\enddata
\tablenotetext{a}{ $\chi^2_r = {1 \over {\nu}} \Sigma_i ((\theta_i - \theta_{Mi})^2 / \sigma_i^2)$  where $\theta_i$ are the data representing the angles of the B-Vector at various locations, i;  $\sigma_i$ is the uncertainty in each angle; $\theta_{Mi}$ is the angle of the B field at the location of each data point for a particular model; and $\nu$ is the number of degrees of freedom for the data set.  $\chi^2_r \sim 1$ for a good fit to the data. }
\tablenotetext{b}{``All Vectors'' means all vectors listed in Table 1 for each source. ``m/Except'' means all vectors listed in Table 1 minus the exceptions discussed in Section 3.2.1. ``WLH03'' means the data shown in Figure 1 of Wolf, Launhardt \& Henning (2003).}
\tablenotetext{c}{For this model comparison with the data, the $\nu$ is one less than the value quoted at the top of each column.}
\tablenotetext{d}{$\theta_{outflow}$ is the angle of the bipolar outflow axis in each source; $\theta_{PDSA}$ is the angle of the symmetry axis  of the pseudo-disk in each source; and $\theta_{PDMA}$ is the angle of the major axis of the pseudo-disk in  each source.  The ALS03b Model is that model shown in Figure 8c of Allen, Li \& Shu (2003).}
\end{deluxetable}

\begin{figure}[p]   
  \begin{center}
 \includegraphics[width=4.5in]{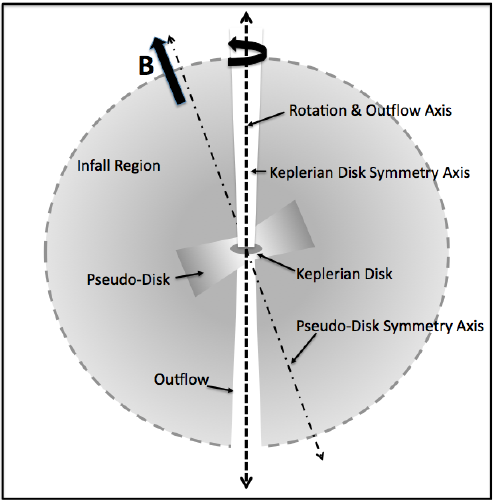} 
  \caption{A cartoon summarizing the magnetically regulated core collapse scenario outlined in Section 1. Typically, the diameter of the infall region is $\sim$10,000 AU, the diameter of a pseudo-disk is $\sim$2,000 AU, and the diameter of a Keplerian disk is $\sim$100 AU. In the magnetically regulated core collapse scenario: the pseudo-disk symmetry axis is aligned with the core magnetic field; and magnetic braking tends to align the core rotation axis with the magnetic field, but this alignment may not be exact.  The pseudo-disk is a dynamically collapsing object formed by the magnetic fields, not rotation.  The Keplerian disk is an object formed by rotation and so its symmetry axis is aligned with the core's rotation axis, as too is the outflow axis if the outflow is driven by rotation. }
   \end{center}
  \label{fig:cartoon}
  \end{figure}

\begin{figure}[p]   
  \begin{center}
 \includegraphics[width=6in]{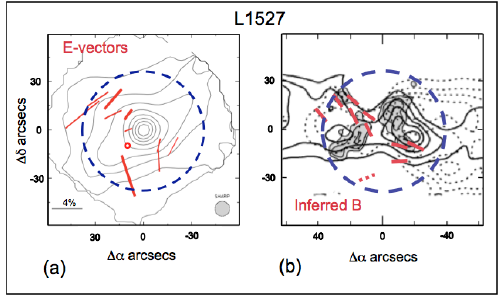} 
  \caption{(a) 350 $\mu$m polarization vectors for L1527.  The thickness of each vector indicates the significance, with all vectors having significance greater than $2\sigma$ before bias correction (see section 2.2). Note the very low polarization ($\le 0.9\% $) marked by a small open circle near the flux peak (see section 2.2).    The contours indicate the total intensity, I, measured by SHARP.  The highest contour level corresponds to 90\% of the peak flux, and each subsequent contour represents a decrement of 10\% of the peak flux.  (b) Inferred magnetic field directions superimposed on a section of a figure from Hogerheijde et al.\ (1998) showing $^{12}$CO 3-2 observations of the bipolar outflow (contours) and a HCO$^+$ map tracing gas on the cavity walls surrounding the outflow (gray area). B-vectors in regions having 350 $\mu$m flux less than 25\% of the peak are shown as dashed lines. The full Hogerheijde et al.\ figure shows the blueshifted outflow (solid contours) extending 2$\arcmin$ east and the redshifted outflow (dashed contours) extending 2$\arcmin$ west from the YSO in L1527. The large dashed circles in (a) and (b) show the extent of the measured infall region for this object (Myers et al.\ 1995; Zhou, Evans \& Wang 1996). }
   \end{center}
  \label{fig:L1527}
  \end{figure}

\begin{figure}[p]
\begin{center}
\includegraphics[width=6in]{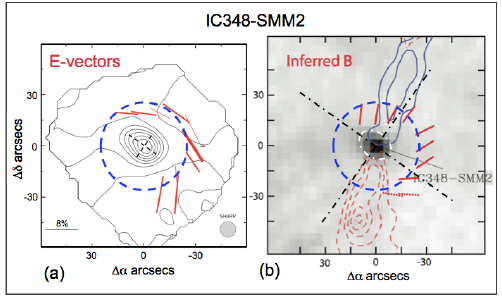}
\caption{\small{
(a) 350 $\mu$m polarization vectors for IC348-SMM2.   The thickness of each vector indicates the significance, with all vectors having significance greater than $2\sigma$ before bias correction (see section 2.2).  The contours indicate the total intensity, I, measured by SHARP.   The highest contour level corresponds to 90\% of the peak flux, and each subsequent contour represents a decrement of 10\% of the peak flux.  The dashed lines crossing at (0,0) represent the major and minor axes of the 65\% contour, which after background subtraction best represents the FWHM of the compact inner core. (b) Inferred magnetic field directions superimposed on a section of a figure from Tafalla et al.\ (2006) which includes a $^{12}$CO 2-1 outflow map (contours) and the 1.2 mm continuum map tracing the dust (grayscale).  B-vectors in regions having 350 $\mu$m flux less than 25\% of the peak are shown as dashed lines.  The full Tafalla et al.\  figure shows the blueshifted outflow (solid contours) extending 1.5$\arcmin$ N-W and the redshifted outflow (dashed contours) extending 1.5$\arcmin$ S-E from IC348-SMM2.  The white dash-dot contour and the black dash-dot straight lines superimposed on this image represent the 65\% intensity contour in (a) and the projection of the major and minor axes of this contour. The large dashed circles in (a) and (b) show the extent of the infall region assumed in this paper for comparison with theory (see section 3.2).}} 
\end{center}
\label{fig:IC348}
\end{figure}

\begin{figure}[p]
\begin{center}
\includegraphics[width=6in]{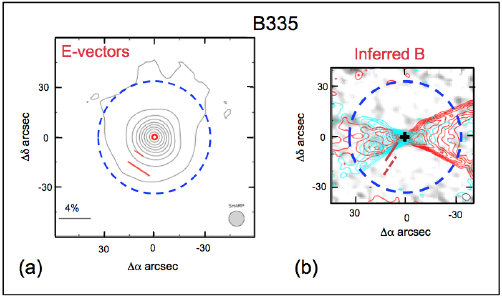}
\caption{\small{
(a) 350 $\mu$m polarimetry vectors for B335. The thickness of each vector indicates the significance, with both vectors having significance greater than $2\sigma$ before bias correction (see section 2.2). Note the very low polarization ($\le 0.6\% $) marked by a small open circle near the flux peak (see section 2.2). The contours indicate the total intensity, I, measured by SHARP.  The highest contour level corresponds to 90\% of the peak flux, and each subsequent contour represents a decrement of 10\% of the peak flux.  (b) Inferred magnetic field directions superimposed on our new interferometric outflow measurements of the $^{12}$CO 1-0 line (contours) using CARMA with a beam size of 4.4$\arcsec \times$ 3.8$\arcsec$. The B-vectors in regions having 350 $\mu$m flux less than 25\% of the peak are shown as dashed lines. The full extent of the outflow as measured by Hirano et al.\ (1988) in $^{12}$CO 1-0 is 4$\arcmin$ east (blue lobe) to 4$\arcmin$ west (red lobe) from the YSO in B335.  The large dashed circles in (a) and (b) show the extent of the measured infall region for this object (Zhou et al.\ 1993). 
}} 
\end{center}
\label{fig:B335}
\end{figure}

\begin{figure}[p]
\begin{center}
\includegraphics[width=6in]{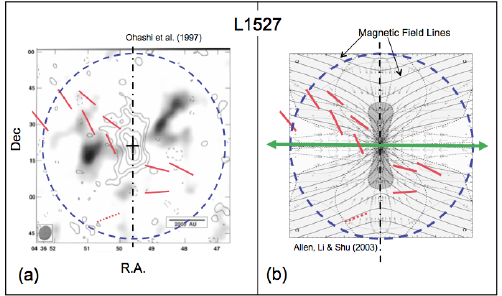}
\caption{\small{
(a) Inferred $B$-vectors (in red) superimposed on interferometric C$^{18}$O (1 - 0) image (contours) of L1527 from Ohashi et al.\ (1997) and $^{12}$CO (1 - 0) redshifted molecular outflow image (grayscale) from Tamura et al. (1996).  (b) Same $B$-vectors superimposed on the core collapse model taken from Figure 8c of Allen, Li \& Shu (2003).  The magnetic field lines in the model are represented by solid lines; the gray area represents the region of highest density (i.e., the pseudo-disk).   The thick green arrow indicates the observed outflow axis, as shown in Figure 2b and discussed in section 3.1.  The model has been aligned with the observed pseudo-disk for L1527, which is shown in (a) and discussed in section 3.1. The straight dash-dot lines in (a) and (b) show the orientation of the major axis of the edge-on pseudo-disk. The large dashed circles in (a) and (b) show the extent of the measured infall region for this object (Myers et al.\ 1995; Zhou, Evans \& Wang 1996). }} 
\end{center}
\label{fig:L1527b}
\end{figure}

\begin{figure}[p]
\begin{center}
\includegraphics[width=6in]{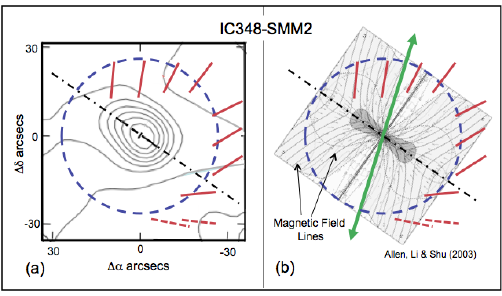}
\caption{\small{
a) Inferred $B$-vectors (in red) superimposed on our 350 $\mu$m intensity image of IC348-SMM2.  (b) Same $B$-vectors superimposed on the core collapse model taken from Figure 8c of Allen, Li \& Shu (2003).  The magnetic field lines in the model are represented by solid lines; the gray area represents the region of highest density (i.e., the pseudo-disk).   The thick green arrow indicates the observed outflow axis, as shown in Figure 3b.  The model has been aligned with the observed pseudo-disk for IC348-SMM2 (see discussion in section 3.1). The straight dash-dot lines in (a) and (b) show the orientation of the major axis of the edge-on pseudo-disk. As in Figure 3, the large dashed circles in (a) and (b) show the extent of the assumed infall region for this object. }}  
\end{center}
\label{fig:IC348b}
\end{figure}

\begin{figure}[p]
\begin{center}
\includegraphics[width=6in]{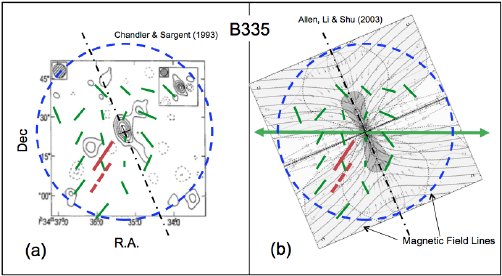}
\caption{\small{
a) Inferred $B$-vectors superimposed on the interferometric 2.7 mm continuum image of B335 from Chandler \& Sargent (1993).  The larger (red) vectors are based on our 350 $\mu$m polarimetry results reported in this paper, the smaller (green) vectors are based on the 850 $\mu$m polarimetry results of Wolf, Launhardt \& Henning (2003).   (b) Same $B$-vectors superimposed on the core collapse model taken from Figure 8c of Allen, Li \& Shu (2003).  The magnetic field lines in the model are represented by solid lines; the gray area represents the region of highest density (i.e., the pseudo-disk).   The thick green arrow indicates the observed outflow axis, as shown in Figure 4b.  The model has been aligned with the observed pseudo-disk for B335, which is shown in (a) and discussed in section 3.1. The straight dash-dot lines in (a) and (b) show the orientation of the major axis of the edge-on pseudo-disk. The large dashed circles in (a) and (b) show the extent of the measured infall region for this object (Zhou et al.\ 1993). }} 
\end{center}
\label{fig:B335b}
\end{figure}

\end{document}